\RequirePackage{fixltx2e}
\documentclass[twoside,journal]{IEEEtran}
\usepackage[T1]{fontenc}
\usepackage[latin9]{inputenc}
\usepackage{textcomp}
\usepackage{graphicx}
\usepackage[unicode=true,pdfusetitle,
 bookmarks=true,bookmarksnumbered=false,bookmarksopen=false,
 breaklinks=false,pdfborder={0 0 1},backref=false,colorlinks=false]
 {hyperref}
\usepackage{breakurl}

\makeatletter
\setkeys{Gin}{width=\linewidth}
\usepackage{cite}

\long\def\@makecaption#1#2{\ifx\@captype\@IEEEtablestring%
\footnotesize\begin{center}{\normalfont\footnotesize #1}\\
{\normalfont\footnotesize\scshape #2}\end{center}%
\@IEEEtablecaptionsepspace
\else
\@IEEEfigurecaptionsepspace
\setbox\@tempboxa\hbox{\normalfont\footnotesize {#1.}~~ #2}%
\ifdim \wd\@tempboxa >\hsize%
\setbox\@tempboxa\hbox{\normalfont\footnotesize {#1.}~~ }%
\parbox[t]{\hsize}{\normalfont\footnotesize \noindent\unhbox\@tempboxa#2}%
\else
\hbox to\hsize{\normalfont\footnotesize\hfil\box\@tempboxa\hfil}\fi\fi}
\ifCLASSOPTIONcompsoc
  \usepackage[caption=false,font=normalsize,labelfont=sf,textfont=sf]{subfig}
\else
  \usepackage[caption=false,font=footnotesize]{subfig}
\fi

\@ifundefined{showcaptionsetup}{}{%
 \PassOptionsToPackage{caption=false}{subfig}}
\usepackage{subfig}
\makeatother

\begin{document}

\title{Anatomical Region-Specific \emph{In Vivo} Wireless Communication
Channel Characterization}

\author{A. Fatih Demir, Student Member, IEEE, Qammer H. Abbasi, Senior Member,
IEEE, Z. Esat Ankarali, Student Member, IEEE, Akram Alomainy, Senior
Member, IEEE, Khalid Qaraqe, Senior Member, IEEE, Erchin Serpedin,
Fellow, IEEE and Huseyin Arslan, Fellow, IEEE\thanks{Manuscript received
June 10, 2016, revised September 09, 2016, and accepted October 13,
2016.} \thanks{Ali Fatih Demir and Z. Esat Ankarali are with the
Department of Electrical Engineering, University of South Florida,
Tampa, FL, USA (e-mail:\{afdemir; zekeriyya\}@mail.usf.edu).} \thanks{
Qammer H. Abbasi, Khalid Qaraqe and Erchin Serpedin are with Department
of ECEN Texas A\&M University, Qatar/USA (email: \{qammer.abbasi,
k.qaraqe, eserpedin\}@tamu.edu).} \thanks{ Akram Alomainy is with
the School of EECS, Queen Mary University of London, London, UK (email:
a.alomainy@qmul.ac.uk).} \thanks{Huseyin Arslan is with the Department
of Electrical Engineering, University of South Florida, Tampa, FL,
USA and also with the College of Engineering, Istanbul Medipol University,
Istanbul, Turkey (e-mail: arslan@usf.edu).}}
\maketitle
\begin{abstract}
\emph{In vivo} wireless body area networks (WBANs) and their associated
technologies are shaping the future of healthcare by providing continuous
health monitoring and noninvasive surgical capabilities, in addition
to remote diagnostic and treatment of diseases. To fully exploit the
potential of such devices, it is necessary to characterize the communication
channel which will help to build reliable and high-performance communication
systems. This paper presents an \emph{in vivo} wireless communication
channel characterization for male torso both numerically and experimentally
(on a human cadaver) considering various organs at 915 MHz and 2.4
GHz. A statistical path loss (PL) model is introduced, and the anatomical
region-specific parameters are provided. It is found that the mean
PL in dB scale exhibits a linear decaying characteristic rather than
an exponential decaying profile inside the body, and the power decay
rate is approximately twice at 2.4 GHz as compared to 915 MHz. Moreover,
the variance of shadowing increases significantly as the \emph{in
vivo} antenna is placed deeper inside the body since the main scatterers
are present in the vicinity of the antenna. Multipath propagation
characteristics are also investigated to facilitate proper waveform
designs in the future wireless healthcare systems, and a root-mean-square
(RMS) delay spread of 2.76 ns is observed at 5 cm depth. Results show
that the \emph{in vivo} channel exhibit different characteristics
than the classical communication channels, and location dependency
is very critical for accurate, reliable, and energy-efficient link
budget calculations.
\end{abstract}


\markboth{IEEE JOURNAL OF BIOMEDICAL AND HEALTH INFORMATICS}{Demir \MakeLowercase{\textit{et al.}}:ANATOMICAL REGION-SPECIFIC \emph{IN VIVO} WIRELESS COMMUNICATION CHANNEL CHARACTERIZATION}

\section{Introduction}

\begin{figure}
\includegraphics{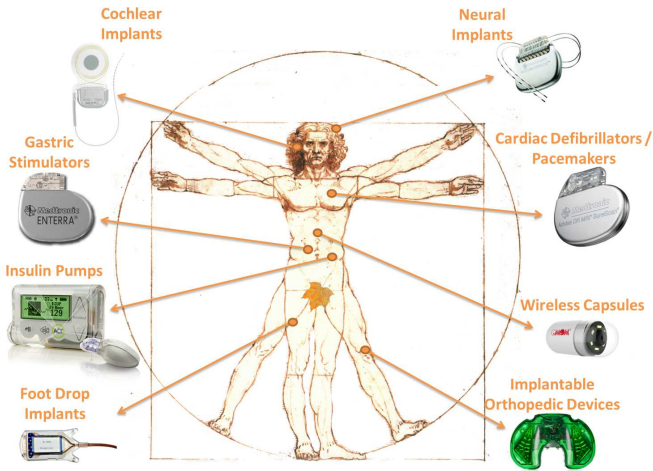}
\centering{}\caption{\emph{In vivo}-WBAN devices for various applications.\label{fig:In-Vivo-WBAN}}
\end{figure}

\IEEEPARstart{C}{hronic} diseases and conditions such as diabetes,
obesity, heart disease, and stroke are the leading causes of death
and disabilities in the United States. Treating people with these
illnesses accounts for 86\%\footnote{http://www.cdc.gov/chronicdisease}
of the national health expenditure which is expected to be almost
double in the next ten years\footnote{https://www.cms.gov}. However,
these are the most preventable and manageable problems among all health
issues by committing to a healthier lifestyle. Continuous health monitoring
helps to achieve this goal by assisting people to engage in their
healthcare and also allows physicians to perform more reliable analysis
by providing the data collected over a large period of time. In addition,
exploitation of this \emph{big data} will replace the traditional
\textquotedblleft one-size-fits-all\textquotedblright{} model with
more personalized healthcare in the near future. Furthermore, noninvasive
surgery and remote treatment are expected to lower the risk of infection,
reduce hospitalization time and accelerate recovery processes. All
these demanding requirements for an effective service quality in healthcare
awakened a general interest in wireless body area networks (WBANs)
research \cite{demir2016invivo,hall2012antennas,abbasi2010onbodyradio,chavez-santiago2015experimental,floor2015inbodyto,anzai2014experimental,kurup2012inbodypath,lin2013characteristics,demir2016bioinspired,demir2016book}.
One component of such advanced technologies is represented by wireless
\emph{in vivo} sensors and actuators, e.g., pacemakers, internal drug
delivery devices, nerve stimulators, and wireless capsules as shown
in Fig.\ref{fig:In-Vivo-WBAN}. \emph{In vivo} medical devices offer
a cost efficient and scalable solution along with the integration
of wearable devices and help to achieve the vision of advanced pervasive
healthcare, anytime and anywhere \cite{demir2016invivo}. Besides
healthcare, the use of \emph{in vivo}-WBANs is also envisioned for
many other applications such as military, athletic training, physical
education, entertainment, safeguarding, and consumer electronics \cite{cao2009enabling,movassaghi2014wireless}.

\emph{In vivo}-WBANs and their associated technologies will shape
the future of healthcare considering all the potentials and the critical
role of these applications. To fully exploit the use of them for practical
applications, it is necessary to obtain accurate channel models that
are mandatory to build reliable, efficient, and high-performance communication
systems. These models are required not only to optimize the quality
of service metrics such as high data rate, low bit-error rate, and
latency but also to ensure the safety of biological tissues by careful
link budget evaluations. Although, on-body wireless communication
channel characteristics have been thoroughly investigated \cite{smith2013propagation,abbasi2010onbodyradio},
the studies on \emph{in vivo} wireless communication channels (implant-to-implant
and implant-to-external device links) are limited. The \emph{in vivo}
channel exhibits different characteristics than those of the more
familiar wireless cellular and Wi-Fi environments since the electromagnetic
(EM) wave propagates through a very lossy environment inside the body,
and dominant scatterers are present in the near-field region of the
antenna. 

The IEEE 802.15.6 standard \cite{2012ieeestandard} was released in
2012 to regulate short-range wireless communications inside or in
the vicinity of the human body. According to this standard, \emph{in
vivo}-WBAN devices operate in the medical device radio communications
service (MedRadio) which uses discrete bands within the 401\textendash 457
MHz spectrum including the previous specification called medical implant
communication service (MICS) band. Despite the fact that MedRadio
bands provide satisfying propagation characteristics inside the human
body \cite{sani2009numerical}, they suffer from lower bandwidths
and larger antenna size issues compared to the antennas designed to
operate at higher frequencies. Therefore, other frequency bands, such
as industrial, scientific, medical (ISM) and ultra-wide band (UWB)
communications bands should also be considered in the upcoming standards
for \emph{in vivo} wireless communications. It is also known that
EM wave propagation inside the human body is strongly related to the
location of the antenna \cite{sani2009numerical,lin2013characteristics}
and hence, the \emph{in vivo} channel should be investigated for a
specific anatomical part. For example, the gastrointestinal tract
has been studied for wireless capsule endoscopy applications \cite{basar2013theuse},
while the heart area has been investigated for implantable cardioverter
defibrillators and pacemakers \cite{sayrafian-pour2010channel}. Although
many \emph{in vivo} path loss (PL) formulas were reported in the literature
\cite{alomainy2009modeling,sayrafian-pour2010channel,khaleghi2011ultrawideband,chavez-santiago2015experimental,floor2015inbodyto,kurup2012inbodypath,anzai2014experimental},
they do not provide location specific PL model parameters to carry
out accurate link budget calculations. Moreover, detailed human body
models are crucial in order to investigate the \emph{in vivo} wireless
communication channel. Various phantoms have been designed to simulate
the dielectric properties of the tissues for numerical and experimental
investigation. The validation of numerical studies with real experimental
measurements is required, however performing experiments on a living
human is strictly regulated. Therefore, physical phantoms \cite{alomainy2009modeling,lin2013characteristics,chavez-santiago2015experimental}
or anesthetized animals \cite{anzai2014experimental,floor2015inbodyto}
are often used for experimental investigations.

This paper presents a numerical and experimental characterization
of the \emph{in vivo} wireless communication channel for male torso
considering various anatomical regions. The location dependent characteristics
of the \emph{in vivo} channel are investigated by performing extensive
simulations at 915 MHz and 2.4 GHz using HFSS\textsuperscript{\textregistered{}}.
A statistical PL formula is introduced, and anatomical region-specific
parameters are provided. The multipath propagation characteristics
of the \emph{in vivo} channel are examined by investigating the polarization
and analyzing the delay spread, which is of particular importance
for broadband applications. In addition to the thorough simulation
studies, experiments are conducted on a human cadaver, and the results
are compared with the numerical studies. The preliminary results were
presented in \cite{demir2014numerical} and \cite{demir2015experimental}.
To the best of authors' knowledge, this is the first study that investigates
the \emph{in vivo} wireless channel for various anatomical regions
both numerically and experimentally on a human cadaver.

The rest of the paper is organized as follows. Section \ref{sec:II}
describes the simulation/experimental setup and explains the measurement
methodology in detail. Section \ref{sec:III} presents the \emph{in
vivo} channel characterization based on the numerical and experimental
investigation. A statistical PL formula is provided along with the
anatomical region-specific parameters, and multipath propagation characteristics
are examined thoroughly. Finally, Section \ref{sec:IV} summarizes
the contributions and concludes the paper.

\begin{figure*}
\subfloat[\label{fig:Male-Torso}]{\includegraphics[bb=-1bp 0bp 699bp 514bp,width=1\columnwidth]{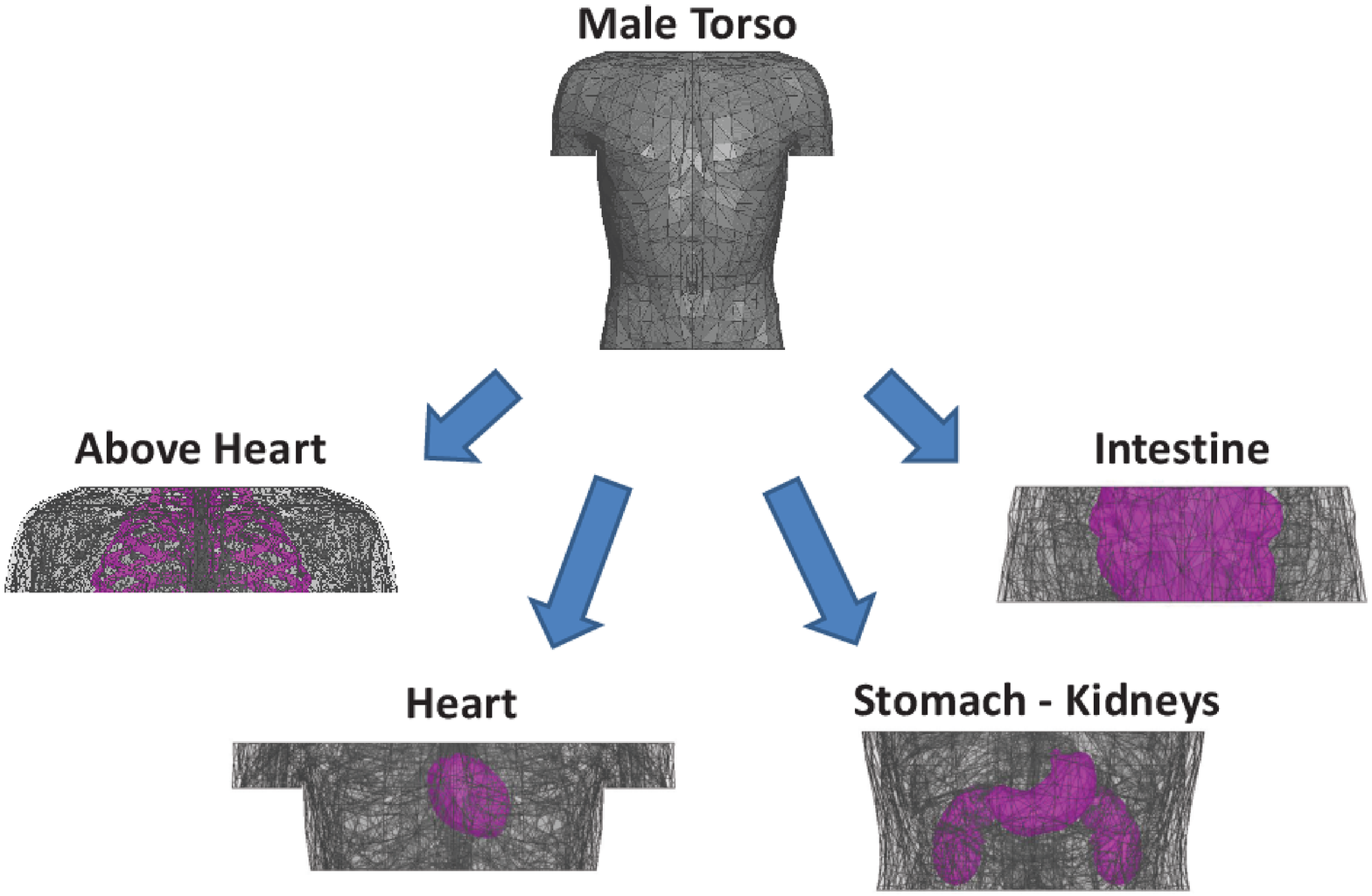}}\subfloat[\label{fig:Sim-Locations}]{\includegraphics[width=1\columnwidth]{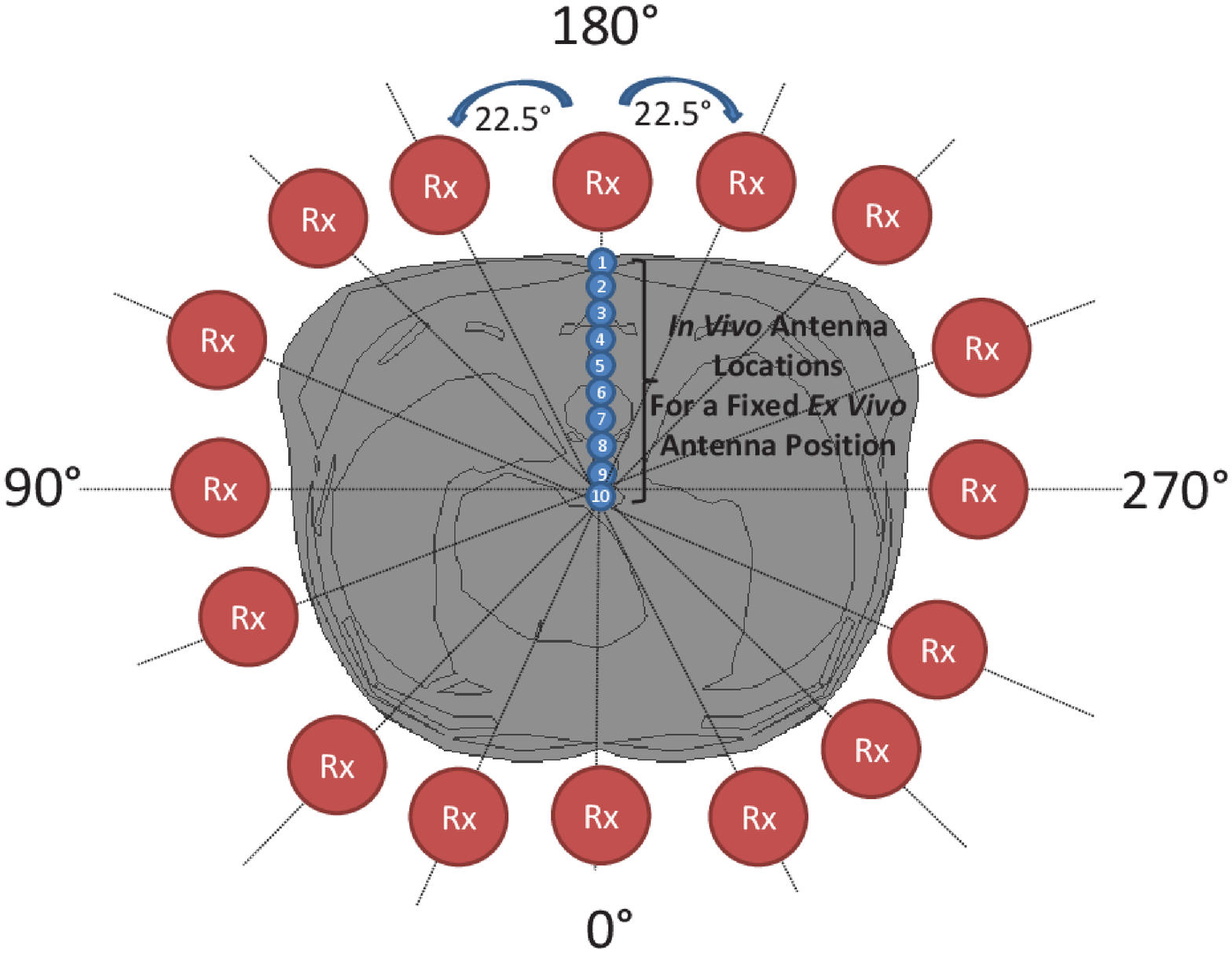}

}

\caption{(a) Investigated anatomical regions; (b) \emph{In vivo} and \emph{ex
vivo} antenna locations in simulations. 16 (angles) x 10 (depth) x
4 (regions) x 2 (operational frequencies) = 1280 simulations were
performed in total for the path loss model.}
\end{figure*}

\section{Simulation and Measurement Settings\label{sec:II}}

\subsection{Simulation Setup}

Analytical methods are viewed as infeasible and require extreme simplifications
\cite{hall2012antennas,pellegrini2013antennas}. Therefore, numerical
methods, which provide less complex and appropriate approximations
to Maxwell\textquoteright s equations, are preferred for characterizing
the \emph{in vivo} wireless communication channel. In this study,
we used ANSYS HFSS\textsuperscript{\textregistered{}} 15.0 \cite{ansyshfss},
which is a full-wave EM field simulator based on the Finite Element
Method (FEM). ANSYS also provides a detailed male human body model,
and it includes frequency dependent dielectric properties of over
300 parts (bones, tissues and organs) with 2 mm resolution. This extensive
simulation work was beyond the capability of personal computers and
advanced computing resources at the University of South Florida (USF)
were used to solve such large EM problems. Research Computing at USF
hosts a computer cluster which currently consists of approximately
500 nodes with nearly 7200 processors cores and 24TB of memory in
total.

The simulations were designed considering an implant to an external
device (in-body to on-body) communications scenario in the male torso
with a similar measurement setup in \cite{demir2014numerical}. Rather
than using the whole body, the torso area was segmented into four
sectors considering the major internal organs: heart, stomach, kidneys,
and intestine as shown in Fig. \ref{fig:Male-Torso}. In each region,
simulations were performed by rotating receiver (\emph{ex vivo}) and
transmitter (\emph{in vivo}) antennas together around the body with
22.5\textdegree{} angle increments (Fig. \ref{fig:Sim-Locations}).
The \emph{ex vivo} antenna was placed 5 cm away from the body surface
and the \emph{in vivo} antenna was placed at ten different depths
(from 10 mm to 100 mm) inside the body for each \emph{ex vivo} antenna
location. In addition, antennas were placed in the same orientation
to avoid polarization losses.

The received power is expressed using the Friss equation (Eq. \ref{eq:Friss})
for free space links\cite{rappaport1996wireless}: 
\begin{eqnarray}
P_{r} & = & P_{t}G_{t}\left(1-\left|S_{11}\right|^{2}\right)G_{r}\left(1-\left|S_{22}\right|^{2}\right)\left(\frac{\lambda}{4\pi R}\right)^{2}\label{eq:Friss}
\end{eqnarray}
where $P_{t}/P_{r}$ represents transmitted/received powers, $G_{t}/G_{r}$
denotes the gain of the transmitter/receiver antenna, $\lambda$ stands
for the free space wavelength, $R$ is the distance between transmitter
and receiver antennas and $\left|S_{11}\right|$, $\left|S_{22}\right|$
are the reflection coefficients of transmitter/receiver antennas.
Unlike free-space communications, \emph{in vivo} antennas are often
considered to be an integral part of the channel \cite{hall2012antennas}
(i.e., the gain cannot be separated from the channel) and hence, they
need to be designed carefully. However, omnidirectional dipole antennas
at 915 MHz and 2.4 GHz were deployed in our simulations for simplicity.
The dipole antenna length is proportional to the wavelength, which
varies with respect to both frequency and permittivity. Higher frequencies
compared to the MedRadio bands provide smaller antenna sizes, hence,
they could be implanted conveniently. In addition, the antennas were
optimized inside the body with respect to the average torso permittivity
for each frequency towards obtaining maximum power delivery. Although
the antennas presented a good return loss (i.e., less than -7dB),
they were perfectly matched by compensating the $\left(1-\left|S_{11}\right|^{2}\right)$
and $\left(1-\left|S_{22}\right|^{2}\right)$ factors to yield a fair
comparison for PL analysis. Also, the mesh size was set to be less
than $\lambda/5$ in this study.

\subsection{Experimental Setup}

\begin{figure}[!b]
\centering\includegraphics[width=1\columnwidth]{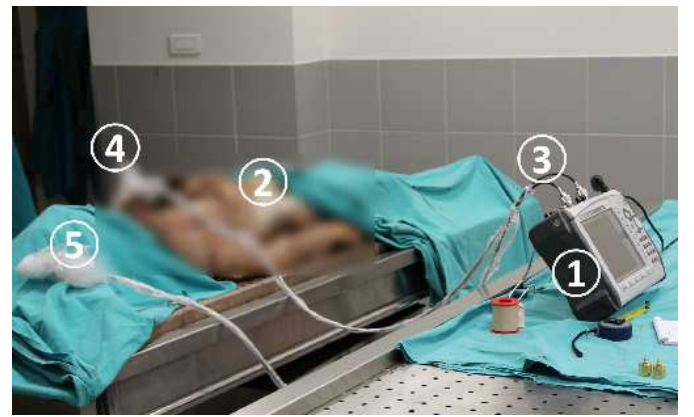}

\caption{Experiment setup for \emph{the in vivo} channel: 1) Vector network
analyzer (VNA), 2) human male cadaver, 3) coaxial cables, 4) a novel
tapered slot CPW-fed antenna (\emph{in vivo}), 5) insulated dipole
antenna (\emph{ex vivo}). \label{fig:Exp-Setup}}
\end{figure}

The numerical investigation was validated by conducting experiments
on a human cadaver in a laboratory environment. Istanbul Medipol University
provided the ethical approval and medical assistance in this study.
The preliminary results were presented in \cite{demir2015experimental}.
Animal organs are used to represent human tissues as suggested in
\cite{stuchly1982dielectric,gabriel1996thedielectric,alomainy2009modeling}
and the decayed human internal organs in this experiment were replaced
with internal organs of a sheep. The male torso area was investigated
at 915 MHz by measuring the channel frequency response, i.e., $S_{21}(f)$,
through a vector network analyzer (VNA). A tapered slot coplanar waveguide
(CPW)-fed antenna \cite{rahman2007anovel} (\emph{in vivo}) and a
dipole antenna (\emph{ex vivo}) were used in our experiments with
two coaxial cables each having a length of 2 meters as illustrated
in Fig. \ref{fig:Exp-Setup}. The frequency response of cables was
subtracted from the channel measurements by performing a calibration
of the VNA. The antennas were wrapped with a biocompatible polyethylene
protective layer and sealed tightly in order not to contact the biological
tissues directly, which could lead to shortening the antennas. The
antennas were tested before the experiment and provided sufficient
return loss inside the body during the experiments (i.e., less than
-7dB).

\begin{figure}
\centering\includegraphics[width=0.6\columnwidth]{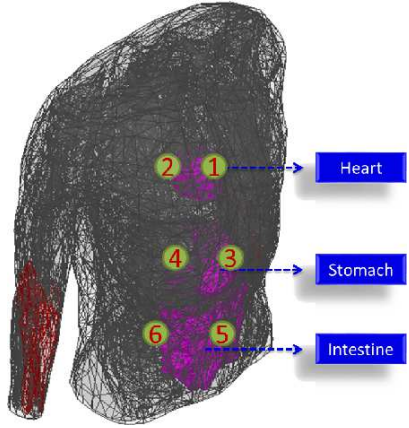}

\caption{Measurement locations on the human cadaver, where odd and even numbers
represent top and bottom of the corresponding organs respectively.
\label{fig:Exp-Locations}}
\end{figure}

The \emph{in vivo} antenna was placed at six different locations (Fig.
\ref{fig:Exp-Locations}) inside the body around the heart, stomach,
and intestine by the help of a physician. \emph{In vivo} depth measurements
were performed using a digital caliper and the antennas were placed
with the same orientation to avoid polarization losses, similar to
the simulations. The channel data was captured between the frequencies
905 MHz to 925 MHz and post processed for further analysis in MATLAB\textsuperscript{\textregistered{}}.
Although the experimental setup did not allow capturing the effects
of circulatory and respiratory systems, it provides a more realistic
multipath propagation scenario than computer simulations or experiments
which are conducted on physical phantoms and anesthetized animals
by providing EM wave propagation in an actual human body.

\begin{figure}[b]
\includegraphics{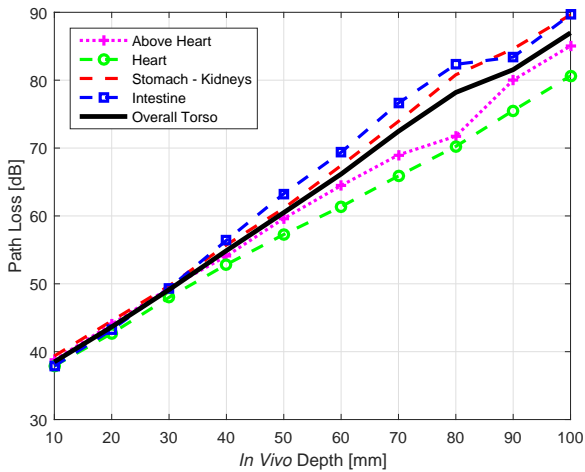}

\caption{Average path loss for four anatomical regions in the simulation environment
at 2.4 GHz. \label{fig:Sim-Depth-PathLoss}}
\end{figure}

\section{\emph{IN VIVO} Channel Characterization\label{sec:III}}

\subsection{Path Loss and Shadowing }

The \emph{in vivo} path loss (PL) expresses a measure of the average
signal power attenuation inside the body and is calculated as $PL=-mean\{|S_{21}|\}$
using the channel frequency response, i.e., $S_{21}$ \cite{floor2015inbodyto,chavez-santiago2015experimental}.
The location dependent characteristic of the \emph{in vivo} PL was
investigated for two ISM bands, i.e., 915 MHz and 2.4 GHz. The EM
wave propagates through various organs and tissues regarding different
antenna locations, and the PL changes significantly even for equal
\emph{in vivo} depths. The location dependent characteristic of the
channel is more dominant when the \emph{in vivo} antenna is placed
deeper inside the body. Fig. \ref{fig:Sim-Depth-PathLoss} presents
the mean PL for the investigated four anatomical body regions in the
simulation environment. Although the signal power attenuation is similar
for near-surface locations, complex body areas such as intestine cause
higher PL due to their dense structure beyond 30 mm \emph{in vivo}
depth. 

\begin{figure}[b]
\includegraphics{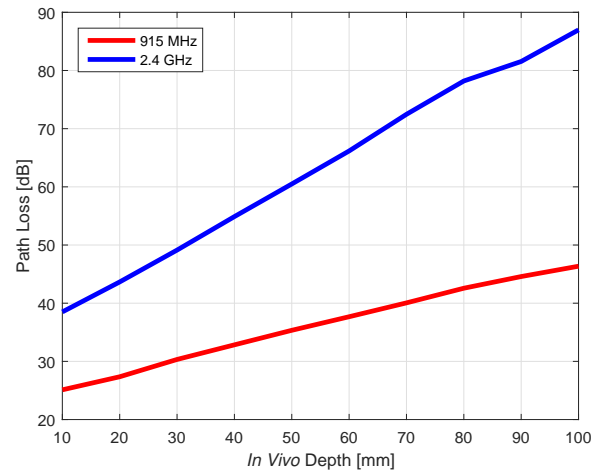}

\caption{Average path loss on torso in the simulation environment at 915 MHz
and 2.4 GHz. \label{fig:Sim-PL-Comp}}
\end{figure}

Various analytical and statistical PL formulas have been proposed
for the \emph{in vivo} channel \cite{demir2016invivo}. Despite the
fact that analytical expressions provide intuition about each component
of the propagation models, they are not practical for link budget
design. According to the final report of the IEEE 802.15.6 standard\textquoteright s
channel modeling subgroup, Friis transmission equation (Eq. \ref{eq:Friss})
can be used for \emph{in vivo} scenarios by adding a random variation
term \cite{Yazdandoost2009Stan,chavez2013propagation}. In this work,
the \emph{in vivo} PL is modeled statistically as a function of depth
by the following equation expressed in dB scale:

\begin{equation}
PL\left(d\right)=PL_{0}+m\left(d/d_{0}\right)+S\qquad\left(d_{o}\leq d\right)\label{eq:2}
\end{equation}
where $d$ represents the depth from the body surface in mm, $d_{0}$
stands for the reference depth with a value of 10 mm, $PL_{0}$ denotes
the intercept term in dB, $m$ is the decay rate of the received power
and $S$ denotes the random shadowing in dB, which presents a normal
distribution for a fixed distance. The power decay rate $(m)$ heavily
depends on the environment and is obtained by performing extensive
simulations and measurements. Also, the shadowing term $(S)$ depends
on the different body materials (e.g., bone, muscle, fat, etc.) and
the antenna gain in different directions \cite{sayrafian-pour2010channel}.
The proposed \emph{in vivo} PL model is valid for $10\le d\le100\,mm$
and the communication channel between an \emph{in vivo} medical device,
and a far external node could be considered as a combination of two
concatenated channels: ``in-body to on-body'' and ``classical indoor
channel'', if there are no surrounding objects around the body \cite{Yazdandoost2009Stan}.
It should be pointed out that the model is antenna dependent as the
majority of other WBAN propagation models in the literature, and this
phenomenon is needed to be taken into account for link budget calculations
as well.

\begin{figure*}
\subfloat[\label{fig:Scatter-915}]{\includegraphics[width=1\columnwidth]{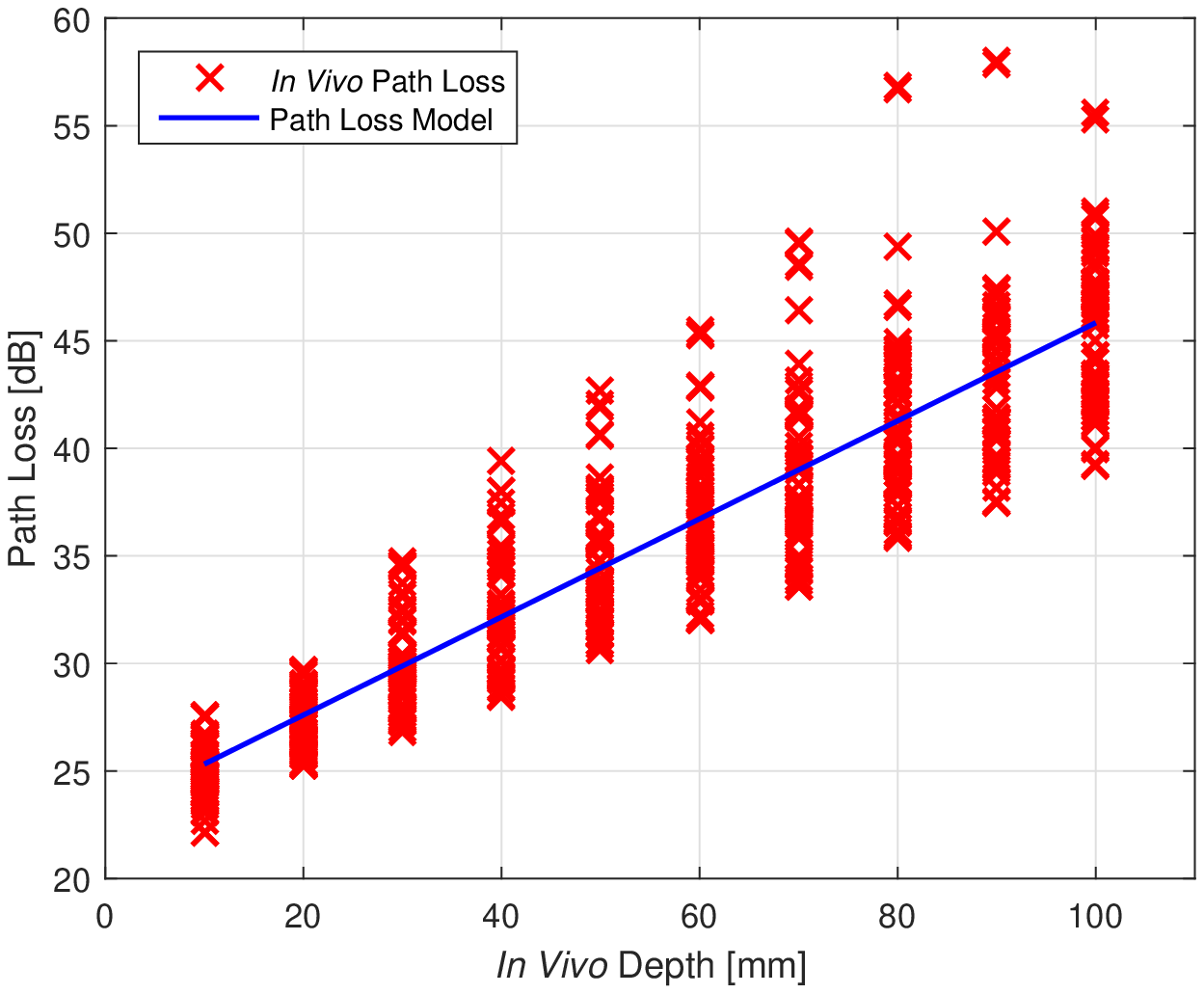}

}\subfloat[\label{fig:Scatter-2400}]{\includegraphics[width=1\columnwidth]{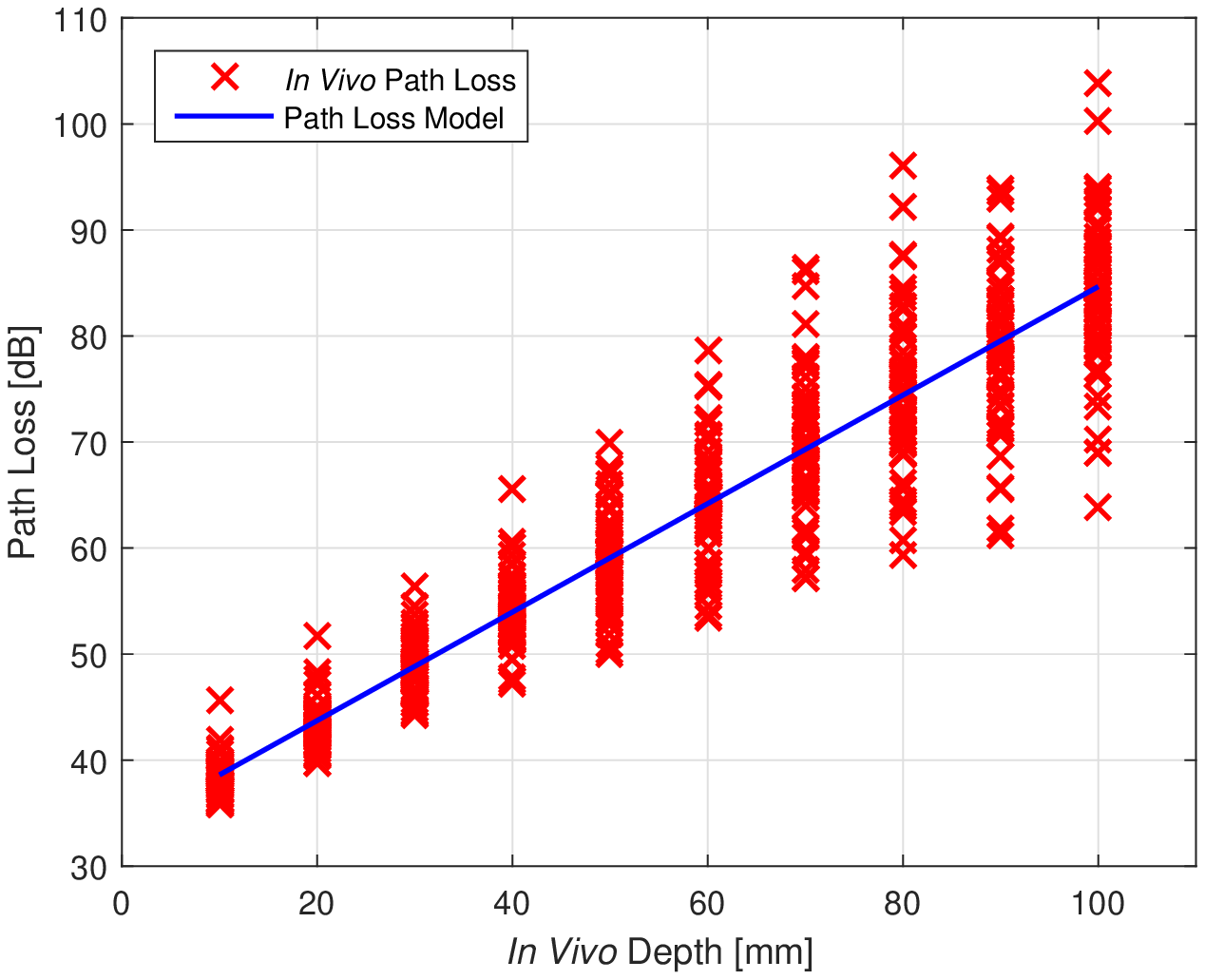}

}

\caption{Scatter plots of path loss vs \emph{in vivo} depth in the simulation
environment at: (a) 915 MHz; (b) 2.4 GHz.\label{fig:Scatter-plots}}
\end{figure*}

\begin{table*}[t]
\caption{Variance Of Shadowing Term $(S)$ In {\upshape dB} For Each In Vivo
Depth\label{tab:Var-of-Shadowing}}

\centering\includegraphics[width=1.15\columnwidth]{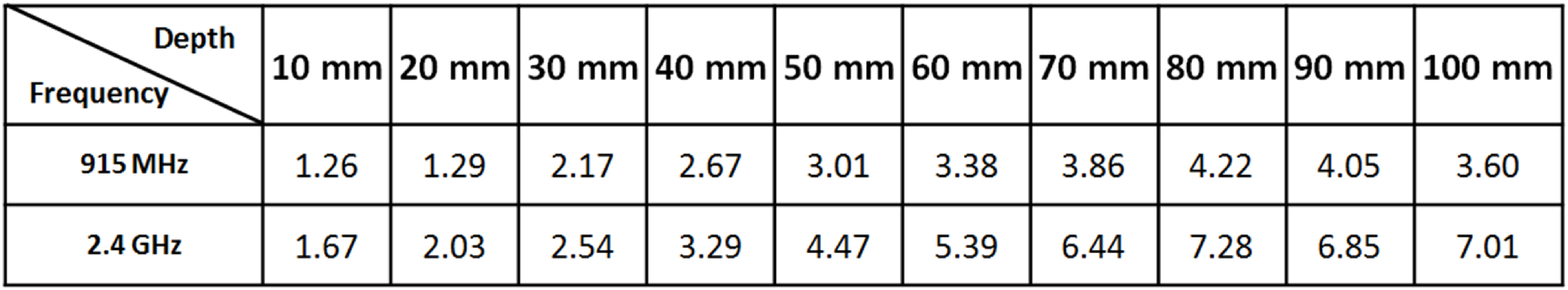}
\end{table*}

\begin{table}[!b]
\caption{Statistical PL model parameters (Anatomical Region) \label{tab:Par-Anat-Region}}

\includegraphics[width=1\columnwidth]{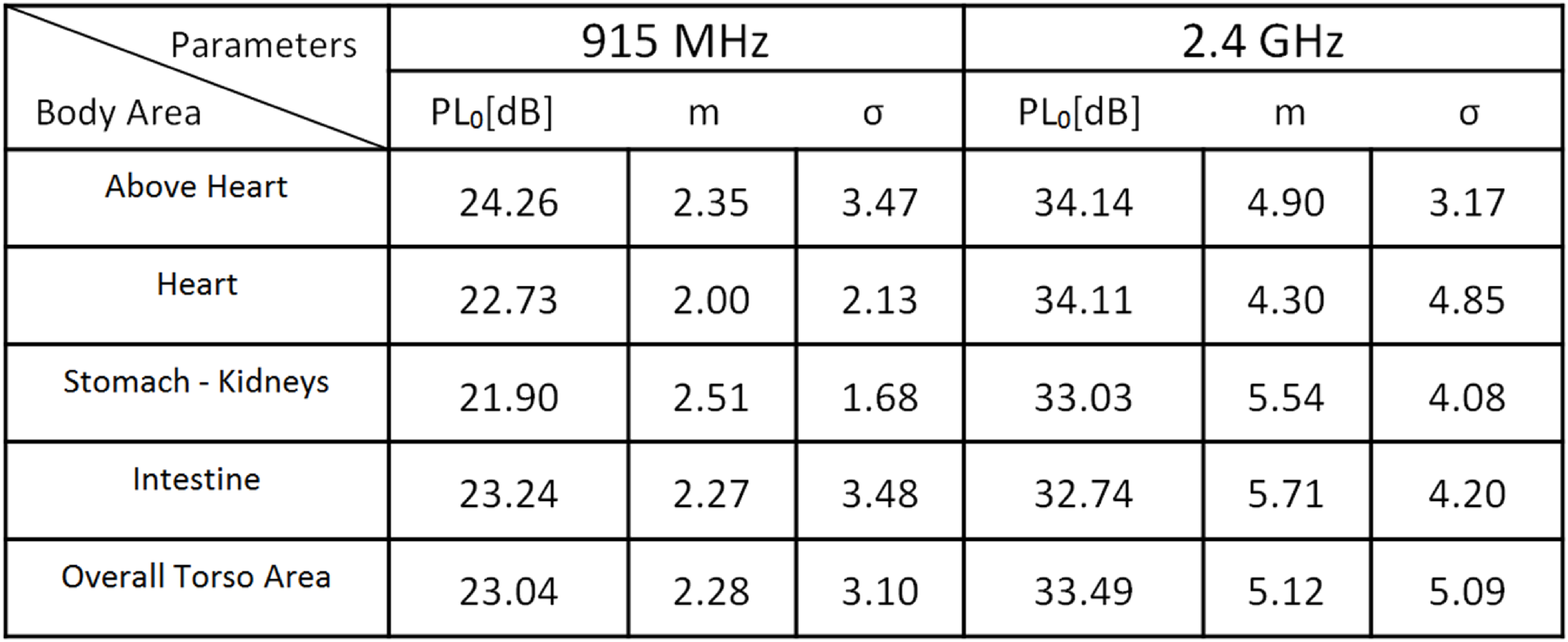}
\end{table}

Figs. \ref{fig:Scatter-plots} shows the scatter plots of PL versus
\emph{in vivo} depth on torso in the simulation environment at 915
MHz and 2.4 GHz. The mean PL is obtained using a linear regression
model. It is observed that the power decay rate $(m)$ is approximately
twice at 2.4 GHz due to the high absorption in tissues as compared
to 915 MHz (Fig. \ref{fig:Sim-PL-Comp}). In addition, the variance
of the shadowing term, $\sigma$, becomes notably larger as the \emph{in
vivo} antenna is placed deeper inside the body as shown in Table \ref{tab:Var-of-Shadowing}.
This behavior can be interpreted using the fundamental statistics
theorem which states that the variance of independent random variables'
sum equals to the sum of the variances of the random variables (scattering
objects) involved in the sum. The \emph{in vivo} channel exhibits
a different characteristic than the classical channels, due to the
main scatterers present in the vicinity of the antenna, and the variance
of shadowing increases significantly compared to free space communications. 

The statistical \emph{in vivo} PL model parameters in Eq. \ref{eq:2}
are provided for each anatomical regions in Tables \ref{tab:Par-Anat-Region}
and \ref{tab:Par-Anat-Dir}, which were obtained by performing extensive
simulations. By interpreting them, it could be concluded that PL increases
when the \emph{in vivo} antenna is placed in an anatomically complex
region. For example, the intestine has a complex structure with repetitive,
curvy-shaped, dissimilar tissue layers, while the stomach exhibits
a smoother structure. As a result, the PL is greater in the intestine
than in the stomach even at equal \emph{in vivo} antenna depths. Also,
more radiation occurs in the posterior region than in the anterior
region due to the human body structure. To sum up, the location dependency
is very critical for link budget calculations and the target anatomical
region should be taken into account to design a high-performance,
energy-efficient communications system inside the body. 

\begin{table}[!b]
\caption{Statistical PL model parameters (Anatomical Direction)\label{tab:Par-Anat-Dir}}

\includegraphics[width=1\columnwidth]{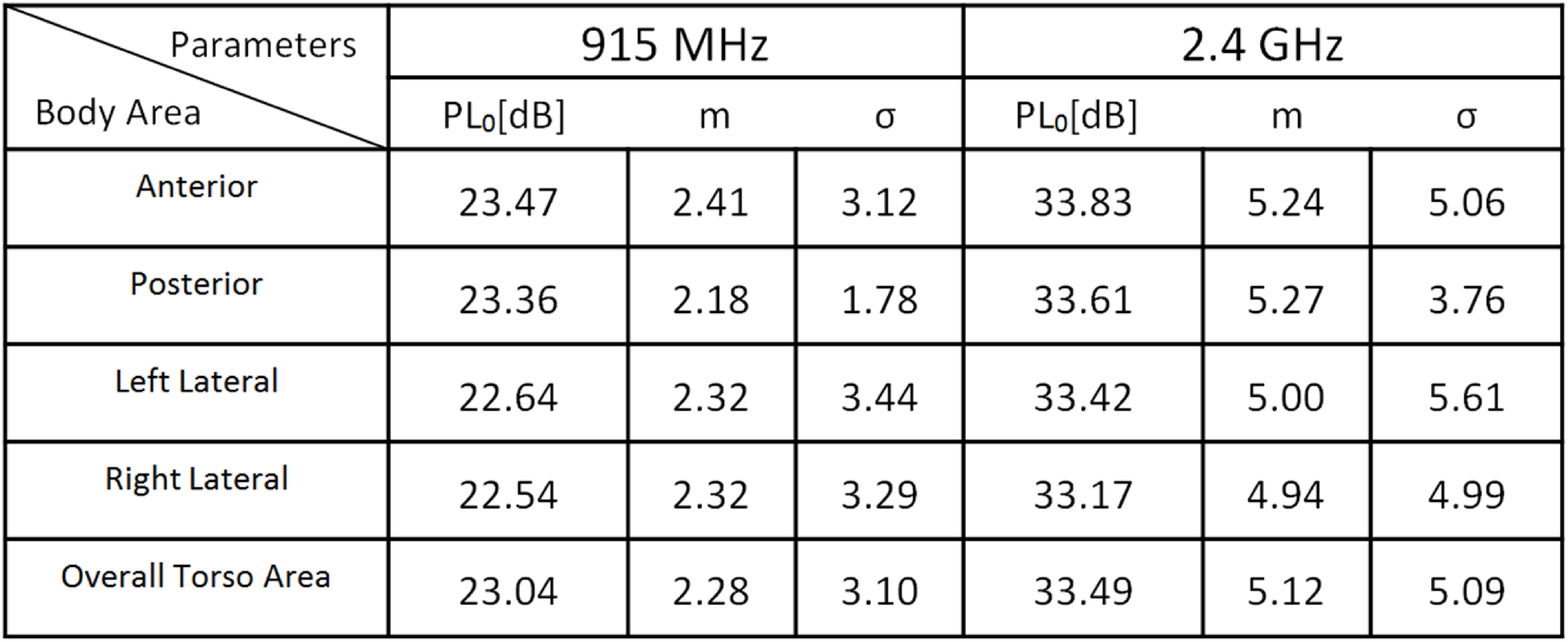}
\end{table}

\begin{table}
\caption{Experimental PL values for selected in vivo locations \label{tab:Exp-PL}}

\centering\includegraphics[width=0.65\columnwidth]{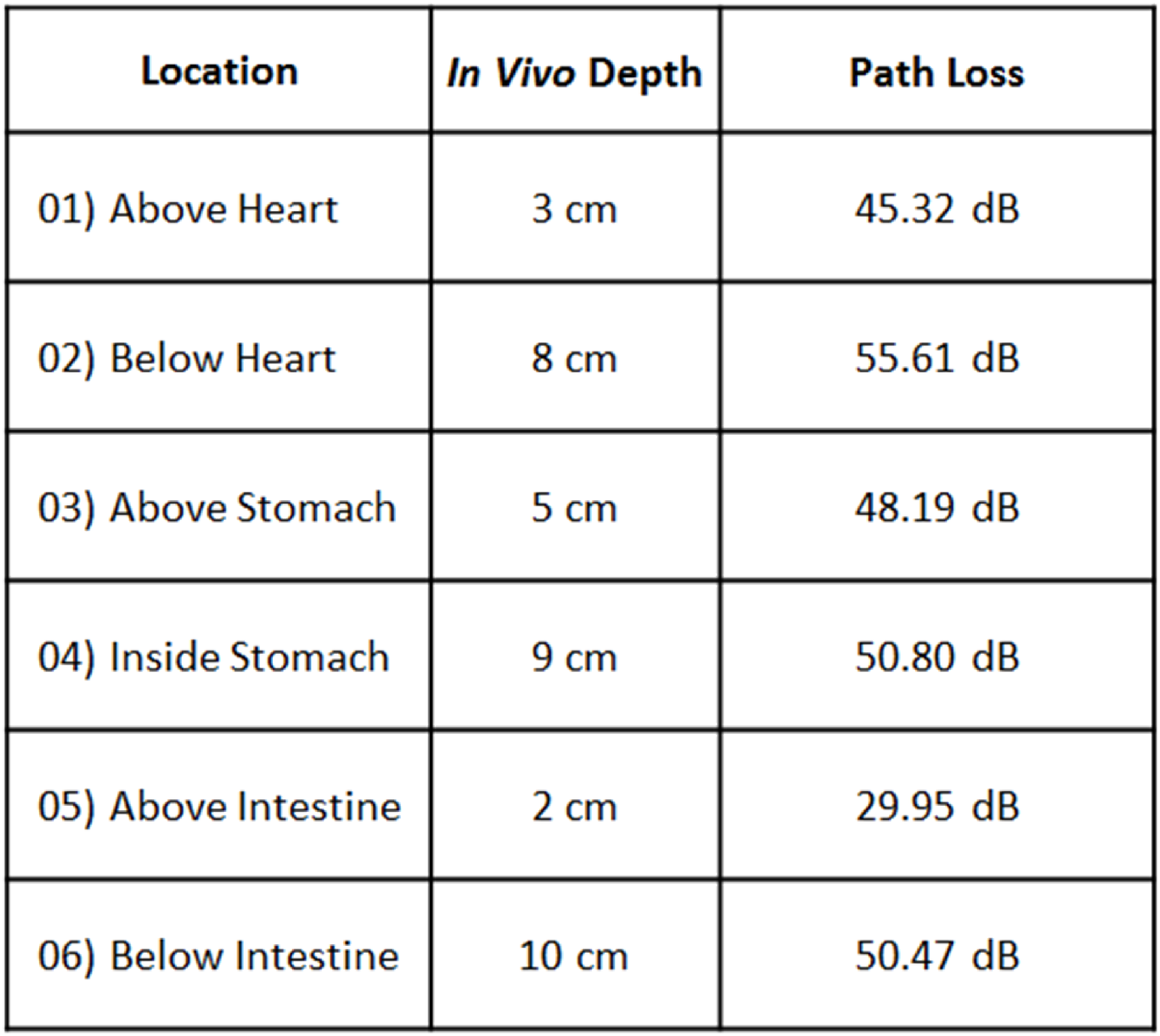}
\end{table}

\begin{figure}[!b]
\centering\includegraphics[width=0.8\columnwidth]{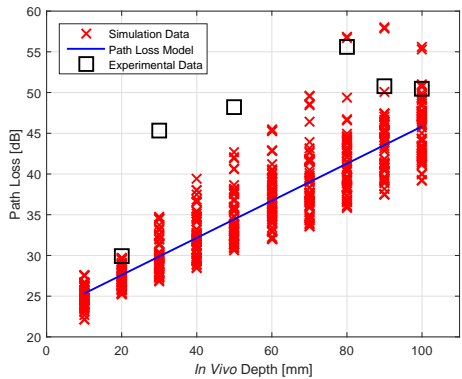}

\caption{Path loss versus \emph{in vivo} depth from the body surface at 915
MHz. \label{fig:Path-loss-versus}}
\end{figure}

The numerical studies were validated with experiments on a human cadaver
at 915 MHz. The \emph{in vivo} antennas were placed at six different
locations as shown in Fig. \ref{fig:Exp-Locations} and the \emph{ex
vivo} antenna was placed 2 cm away from the body surface. Table \ref{tab:Exp-PL}
presents the PL values for the selected \emph{in vivo} locations and
comparison of experimental results with numerical studies are provided
in Fig. \ref{fig:Path-loss-versus}. The discrepancies between the
simulated and measured results exist due to the additional losses
(e.g., antenna distortion), which were not considered in the simulations
and the differences in experimental environment. The statistical \emph{in
vivo} PL model parameters are also provided for the experimental study
and compared with the numerical study in Table \ref{tab:Comp-Sim-Exp}.

\subsection{Multipath Characteristics }

\begin{figure}[b]
\centering\includegraphics[width=0.9\columnwidth]{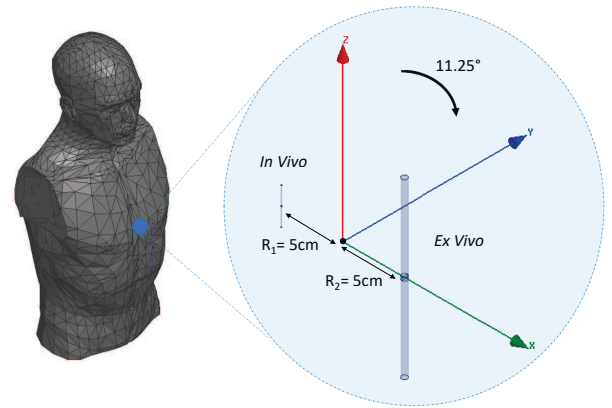}

\caption{Simulation setup for the \emph{in vivo} polarization investigation.
\label{fig:Pol-Setup}}

\end{figure}

\begin{figure}
\includegraphics{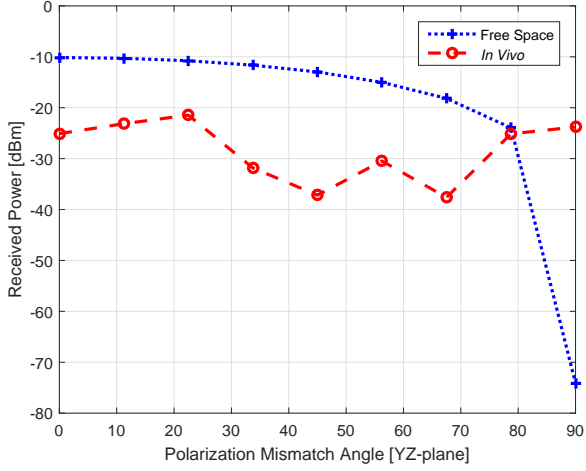}

\caption{Received power for various polarization mismatch angles in the simulation
environment at 915 MHz. \label{fig:Pol-Results} }

\end{figure}

In addition to the PL and shadowing, multipath propagation characteristics
of the \emph{in vivo} channel are also important and should be investigated
to discuss proper waveform designs. Received signal strength was explored
for various antenna polarizations towards understanding the existence
of multipath reflections in the human body medium. As similar to the
previous part, the dipole antennas at 915 MHz were deployed in the
simulation environment, and they were perfectly matched as mentioned
in Section \ref{sec:II}. The \emph{in vivo} antenna was placed at
5 cm depth on the chest, and the \emph{ex vivo} antenna was placed
5 cm away from the body surface to investigate the \textquotedblleft in-body
to off-body\textquotedblright{} link. As a baseline to compare with
the \emph{in vivo} channel, the antennas were separated from each
other by 10 cm in free space. The \emph{ex vivo} antenna was rotated
with 11.25\textsuperscript{o} increments in the YZ-plane for both
scenarios as shown in Fig. \ref{fig:Pol-Setup} and the maximum available
power at the receiver for different polarization mismatch angles is
presented in Fig. \ref{fig:Pol-Results} . In the free space link,
the received power degrades dramatically as the polarization mismatch
increases due to the absence of multipath components, i.e., only line-of-sight
components are effective on the received signal strength. On the other
hand, the received signal power does not change significantly with
polarization mismatch for \emph{in vivo} medium. Therefore, it can
be concluded that biological tissues inside the human body do not
absorb the EM waves completely at 915 MHz and allow reflections that
lead to multipath propagation. These reflections will cause small-scale
fading which is defined as variations over short distances due to
constructive and destructive additions of the signals.

\begin{table}[!t]
\caption{Comparison of the statistical PL model parameters\label{tab:Comp-Sim-Exp}}

\centering\includegraphics[width=0.5\columnwidth]{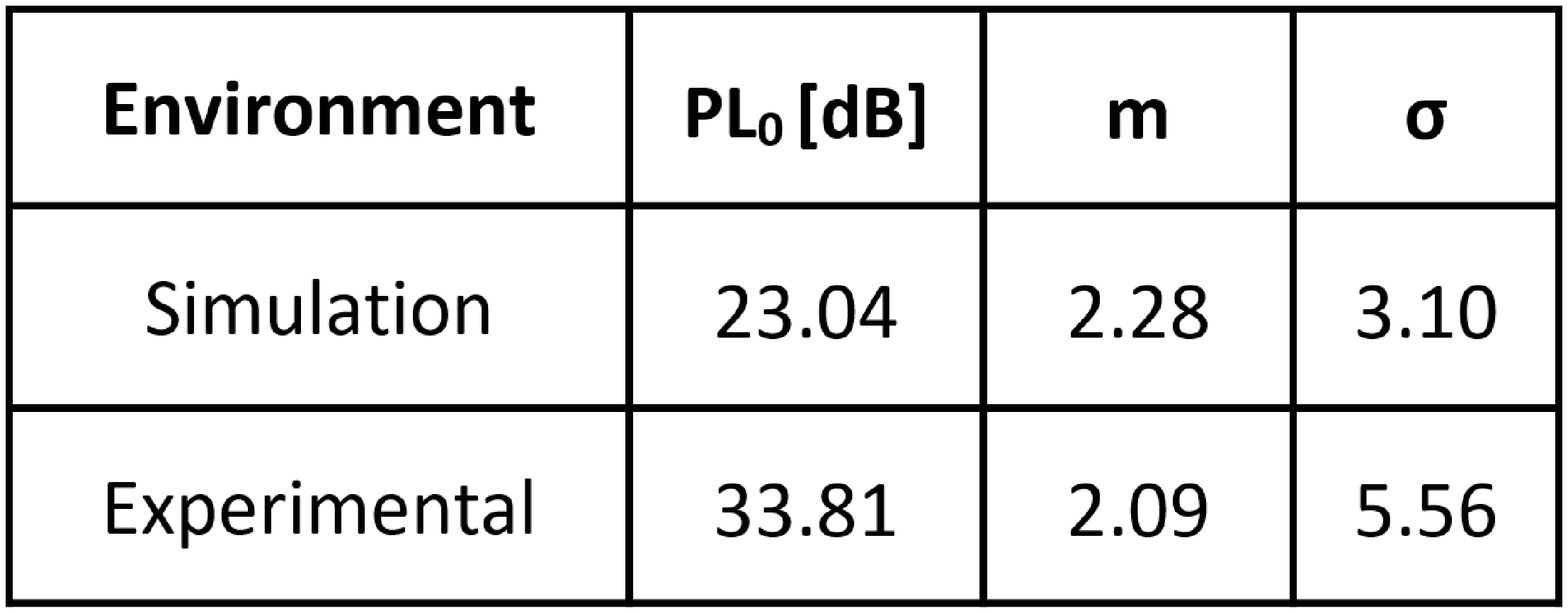}
\end{table}

\begin{figure}[!b]
\centering\includegraphics[width=0.75\columnwidth]{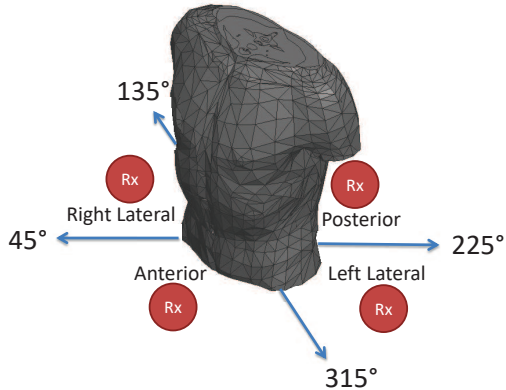}

\caption{Simulation setup for the \emph{in vivo} delay spread investigation.\label{fig:PDP-Setup}}
\end{figure}

\begin{figure}[!t]
\includegraphics{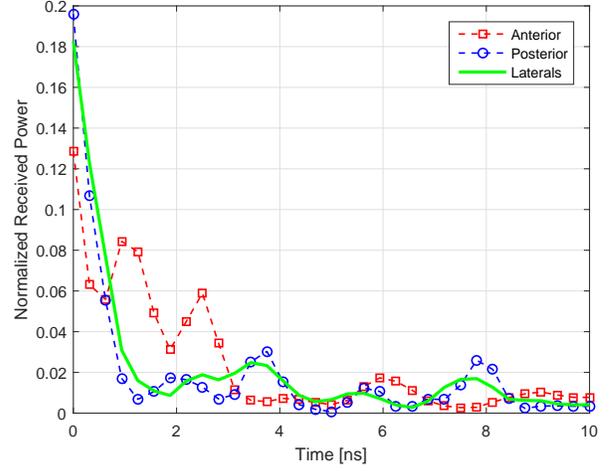}

\caption{Power delay profiles for each anatomical direction in the simulation
environment at 915 MHz. \label{fig:PDP}}
\end{figure}

As a result of multipath propagation inside the human body, the amount
of delay spread should be understood to design an efficient \emph{in
vivo} communications systems. Therefore, power delay profiles (PDPs)
for various anatomical regions were extracted from the simulation
results. The \emph{in vivo} antennas were placed at 5 cm depth on
the torso, and the \emph{ex vivo} antennas were placed 5 cm away from
the body surface as shown in Fig. \ref{fig:PDP-Setup} for four different
directions at 915 MHz. The channel impulse response, $h(t)$, was
obtained by taking the inverse discrete Fourier transform (IDFT) of
the channel frequency response, $S_{21}$. The PDP was calculated
as $PDP(t)=|h(t)|^{2}$ and the total power is normalized to 1 as
presented in Fig. \ref{fig:PDP}. Related multipath channel statistics,\emph{
mean excess delay }($\overline{\tau}$), and \emph{RMS delay spread}
($\sigma_{\tau}$) are calculated to quantify the time dispersion
effect of the \emph{in vivo} channel as follows\cite{lee1988mobilecellular}:

\begin{equation}
\overline{\tau}=\frac{{\displaystyle \sum_{i}}\tau_{i}P(\tau_{i})}{{\displaystyle \sum_{i}}P(\tau_{i})}\label{eq:3}
\end{equation}

\begin{equation}
\sigma_{\tau}=\sqrt{\overline{\tau^{2}}-(\overline{\tau})^{2}}=\sqrt{\frac{{\displaystyle \sum_{i}}\tau_{i}^{2}P(\tau_{i})}{{\displaystyle \sum_{i}}P(\tau_{i})}-\left(\frac{{\displaystyle \sum_{i}}\tau_{i}P(\tau_{i})}{{\displaystyle \sum_{i}}P(\tau_{i})}\right)^{2}}\label{eq:4}
\end{equation}
where $P(\cdot)$ represents the received power in linear scale and,
$\tau_{i}$ denotes the arrival time of the $i^{th}$ path. These
parameters for various anatomical directions are listed in Table \ref{tab:PDP-Table}
and it is observed that the maximum difference in $\sigma_{\tau}$
is 0.3 ns. Therefore, it can be stated that there is almost no difference
in delay spread for various locations when the antennas are implanted
with 5 cm depth on the torso. 

RMS delay spread determines the coherence bandwidth ($B_{c}$) of
the channel. It is a statistical measure of the range of frequencies
where the channel can be assumed as \textquotedblleft flat\textquotedblright{}
\cite{rappaport1996wireless} and the 90\% $B_{c}$ is estimated as
follows: 

\begin{equation}
B_{c}\approx\frac{1}{50\sigma_{\tau}}\label{eq:5}
\end{equation}
The average $\sigma_{\tau}$ at 5 cm \emph{in vivo} depth is measured
as 2.76 ns on the torso and 7.25 MHz coherence bandwidth was predicted
using Eq. \ref{eq:5}. Theoretically, inter-symbol interference (ISI)
is not a critical problem when the signal bandwidth (BW) is less than
$B_{c}$. Hence, the measured delay spread should not cause serious
ISI for narrow-band (NB) communications. However, this dispersion
may lead to a significant interference for UWB communications, which
occupies a BW of greater than 500 MHz. 

\begin{table}[b]
\caption{In vivo multipath propagation statistics at 915 MHz \label{tab:PDP-Table}}
\centering\includegraphics[width=0.55\columnwidth]{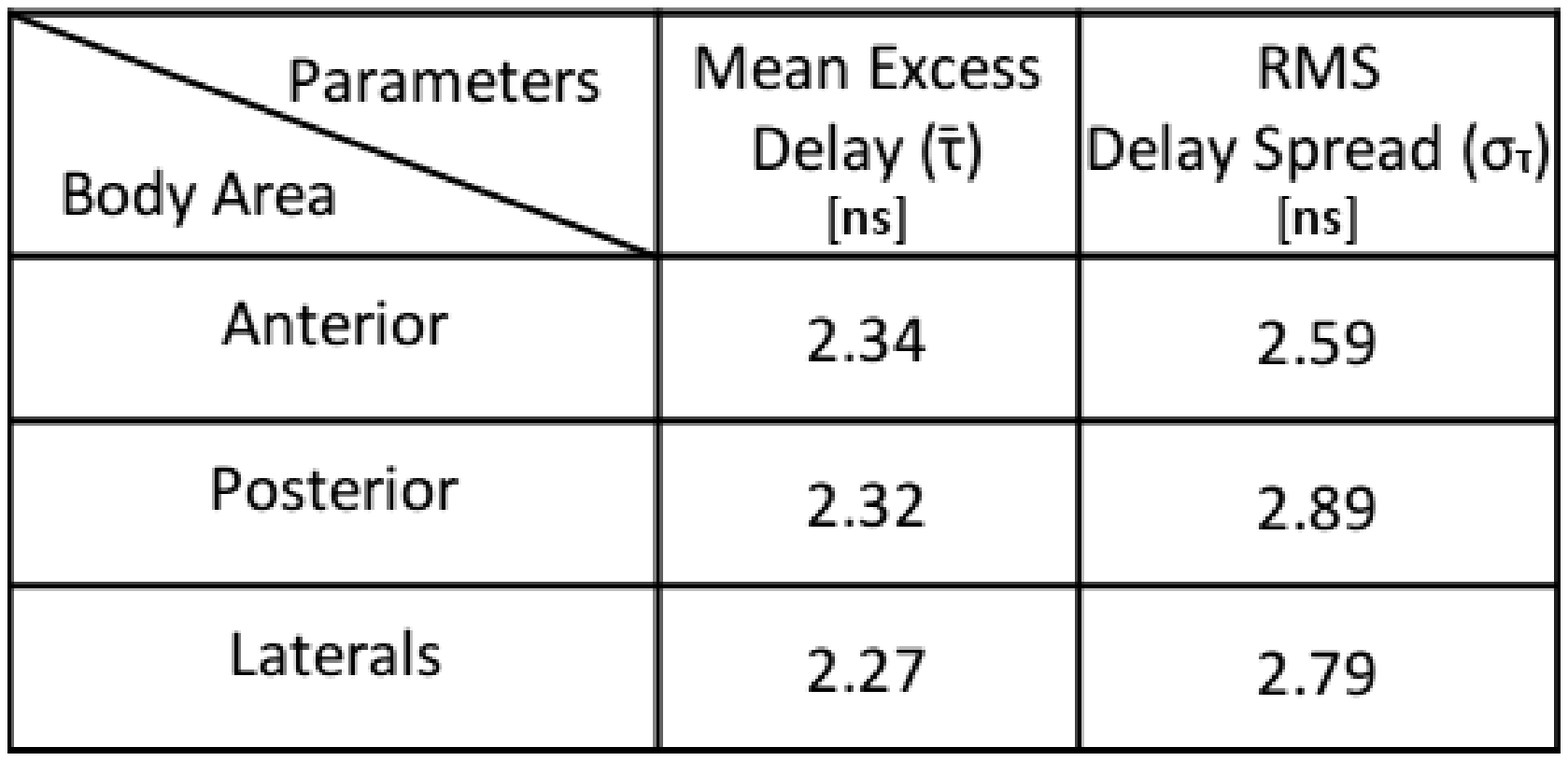}
\end{table}

In frequency-selective channels (i.e., the signal BW is greater than
$B_{c}$) single-carrier waveforms might not exhibit a sufficient
bit error rate (BER) performance without deploying complex equalizers
to solve the ISI problem. Nevertheless, power limitation is a major
constraint for \emph{in vivo}-WBAN devices and hence, the complexity
of signal processing operations must be low. Multi-carrier systems
are offered to provide a trivial solution for the ISI problem. For
example, orthogonal frequency division multiplexing (OFDM) based waveforms
can easily handle delay spread using a cyclic prefix. However, high
peak-to-average-power ratio (PAPR) emerges as a common problem in
multi-carrier waveforms, and it makes the signal vulnerable against
the non-linear characteristics of the radio frequency (RF) front-end
components. Since the \emph{in vivo}-WBAN devices are restricted in
size, the use of high-quality components with high dynamic ranges
is impractical. Therefore, PAPR remains as an important issue and
may still lead the designers to use single-carrier signaling techniques.
To sum up, there are tradeoffs in waveform selection considering the
dispersive nature of the \emph{in vivo} channel and practical issues
together. The system requirements in terms of throughput, power efficiency,
and signal quality need to be clearly identified, and the most proper
waveform technology should be selected accordingly. 

\section{Conclusions\label{sec:IV}}

This paper presented the location dependent characteristics of the
\emph{in vivo} wireless communications channel for male torso at 915
MHz and 2.4 GHz. Extensive simulations were performed using a detailed
3D human body model and measurements were conducted on a human cadaver.
A statistical \emph{in vivo} path loss model is introduced along with
the anatomical region-specific parameters. It is observed that the
path loss in dB scale follows a linear decaying profile instead of
an exponential characteristic, and the power decay rate is approximately
twice at 2.4 GHz as compared to 915 MHz. In addition, the variance
of shadowing increases significantly as the \emph{in vivo} antenna
is placed deeper inside the body since the main scatterers are present
in the vicinity of the antenna. Results show that the location dependency
is very critical for link budget calculations, and the target anatomical
region should be taken into account to design a high-performance \emph{in
vivo} communications system without harming the biological tissues.
Multipath propagation characteristics are examined as well to facilitate
proper waveforms inside the body by investigating various antenna
polarizations and PDPs. A mean RMS delay spread of 2.76 ns is observed
at 5 cm \emph{in vivo} depth. Despite the fact that this dispersion
may not cause significant ISI for NB communications, it could be a
serious issue for UWB communications. The interest in WBANs is rapidly
growing and \emph{in vivo} medical devices are shaping the future
of healthcare. This study will contribute significantly to the upcoming
WBAN standards and hence, will lead to the design of better \emph{in
vivo} transmitter/receiver algorithms.

\section*{ACKNOWLEDGMENT}

This publication was made possible by NPRP grant \# NPRP 6 - 415 -
3 - 111 from the Qatar National Research Fund (a member of Qatar Foundation).
The authors are thankful to Istanbul Medipol University, School of
Medicine for providing the human cadaver and their valuable medical
assistance.

\bibliographystyle{IEEEtran}
\bibliography{BCC_Ref}
\begin{IEEEbiography}[\includegraphics{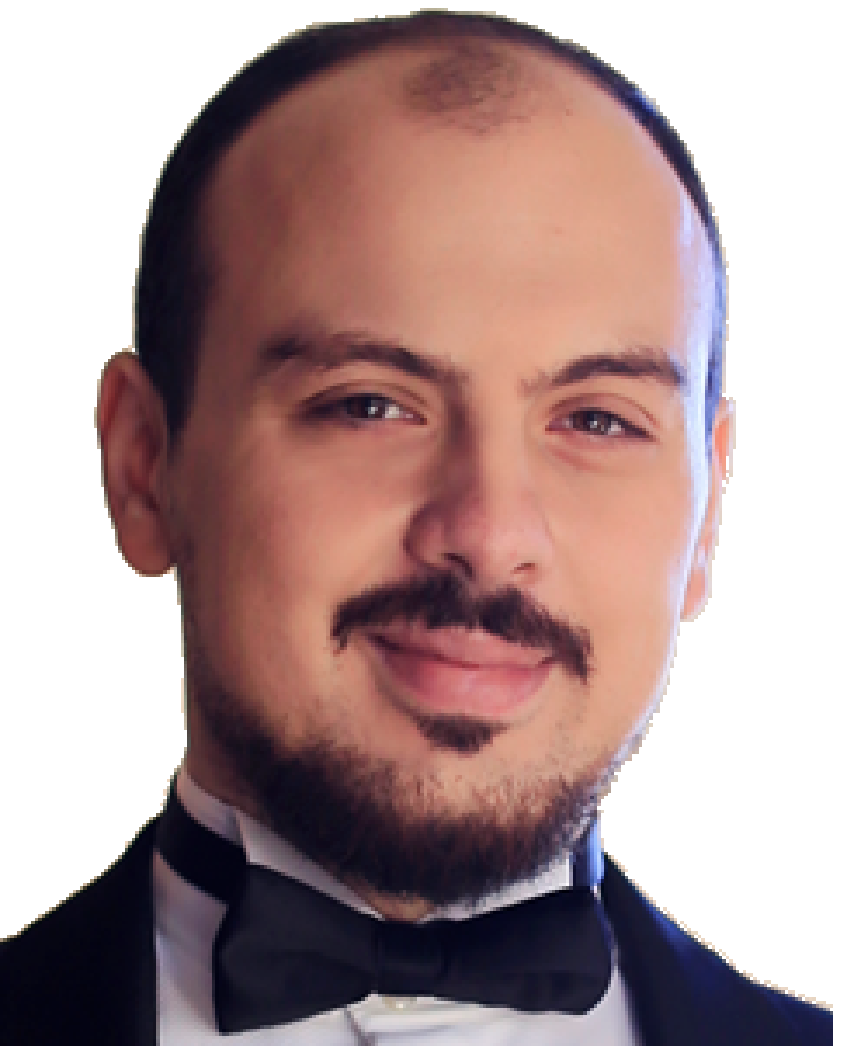}]{Ali Fatih Demir}
(S'08) received his B.S. degree in electrical engineering from Yildiz
Technical University, Istanbul, Turkey, in 2011 and his M.S. degrees
in electrical engineering and applied statistics from Syracuse University,
Syracuse, NY, USA in 2013. He is currently pursuing his Ph.D. degree
as a member of the Wireless Communication and Signal Processing (WCSP)
Group in the Department of Electrical Engineering, University of South
Florida, Tampa, FL, USA. His current research interests include \emph{in
vivo} wireless communications, biomedical signal processing, and brain-computer
interfaces. He is a student member of the IEEE.
\end{IEEEbiography}

\begin{IEEEbiography}[\includegraphics{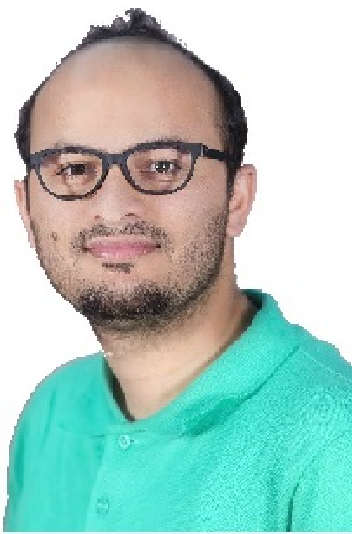}]{Qammer H. Abbasi}
 (S'08\textendash M'12\textendash SM'16) received his B.S. degree
in electronics engineering from the University of Engineering and
Technology, Lahore, Pakistan in 2007, his Ph.D. degree in electronic
and electrical engineering from Queen Mary University of London, U.K.,
in 2012 where he has been a visiting research Fellow since 2013. He
joined the Department of Electrical and Computer Engineering (ECEN)
of Texas A\&M University at Qatar in August 2013, where he is now
an assistant research scientist. His research interests include compact
antenna design, radio propagation, body-centric wireless communications,
cognitive/cooperative network, and MIMO systems. He is a member of
IET and a senior member of the IEEE.
\end{IEEEbiography}

\begin{IEEEbiography}[\includegraphics{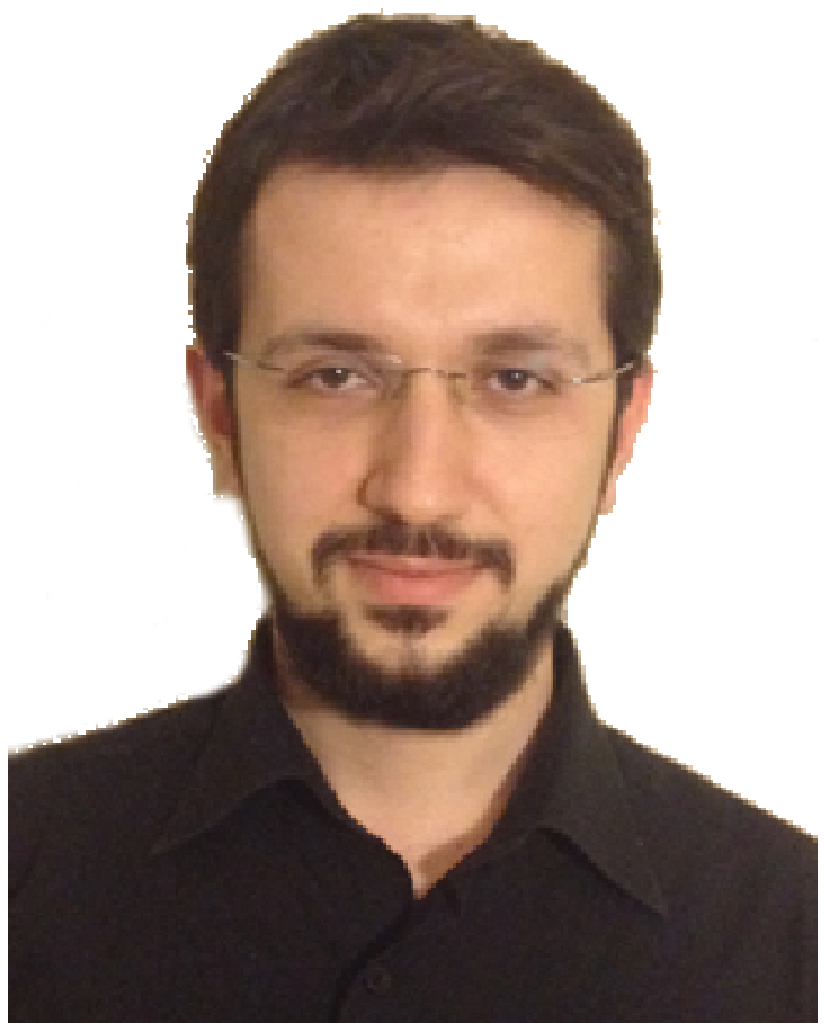}]{Z. Esat Ankarali}
(S'15) received his B.S. degree in control engineering from Istanbul
Technical University, Istanbul, Turkey, in 2011 and his M.S. degree
in electrical engineering from the University of South Florida, Tampa,
FL, USA, in 2013, where he is currently pursuing a Ph.D. degree as
a member of the WCSP Group in the Department of Electrical Engineering.
His current research interests include multi-carrier systems, physical
layer security, and \emph{in vivo} wireless communications. He is
a student member of the IEEE.
\end{IEEEbiography}

\begin{IEEEbiography}[\includegraphics{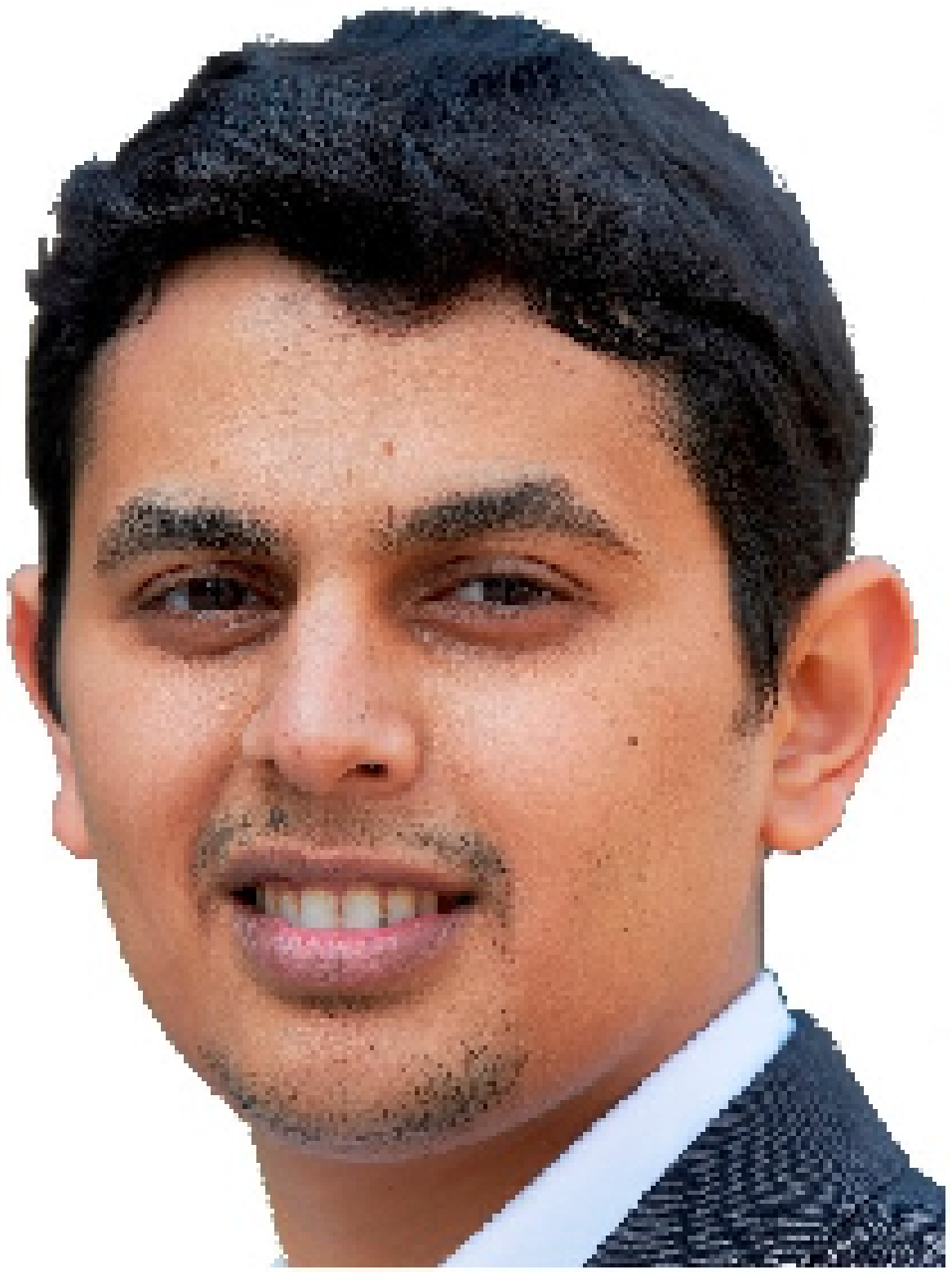}]{Akram Alomainy}
 (S'04\textendash M'07\textendash SM'13) received the M.Eng. degree
in communication engineering and the Ph.D. degree in electrical and
electronic engineering from Queen Mary University of London (QMUL),
U.K., in 2003 and 2007, respectively. He joined the School of Electronic
Engineering and Computer Science, QMUL, in 2007, where he is an associate
professor with the Antennas and Electromagnetics Research Group. His
research interests include compact antenna design for WBANs, radio
propagation characterization, antenna interactions with human body,
and advanced algorithms for intelligent antenna systems. He is a member
of the IET, and a senior member of the IEEE. 
\end{IEEEbiography}

\begin{IEEEbiography}[\includegraphics{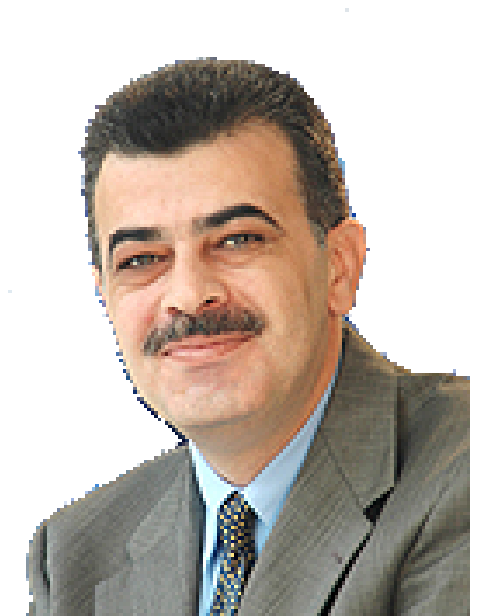}]{Khalid Qaraqe}
 (M\textquoteright 97\textendash SM\textquoteright 00) received
his B.S. degree in electrical engineering (EE) from the University
of Technology, Baghdad, Iraq, in 1986. He received his M.S. degree
in EE from the University of Jordan, Amman, in 1989, and he received
his Ph.D. degree in EE from Texas A\&M University, College Station,
TX, USA, in 1997. He joined the Department of ECEN of Texas A\&M University
at Qatar in July 2004, where he is now a professor. His research interests
include communication theory, mobile networks, cognitive radio, diversity
techniques, and beyond fourth-generation systems. He is a senior member
of the IEEE.
\end{IEEEbiography}

\begin{IEEEbiography}[\includegraphics{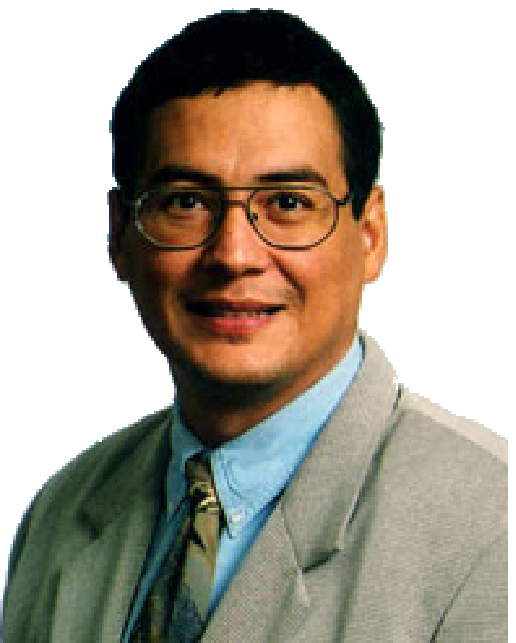}]{Erchin Serpedin}
 (S\textquoteright 96\textendash M\textquoteright 99\textendash SM\textquoteright 04\textendash F\textquoteright 13)
received his specialization degree in signal processing and transmission
of information from Ecole Superieure D\textquoteright Electricite,
Paris, France, in 1992, his M.S. degree from the Georgia Institute
of Technology, Atlanta, GA, USA in 1992, and his Ph.D. degree in electrical
engineering from the University of Virginia, Charlottesville, VA,
USA, in January 1999. He is a professor with the Department of ECEN,
Texas A\&M University, College Station, TX, USA. He is serving as
an associate editor of IEEE Signal Processing Magazine and as the
editor-in-chief of European Association for Signal Processing Journal
on Bioinformatics and Systems Biology. His research interests include
signal processing, biomedical engineering, and machine learning. He
is a fellow of the IEEE. 
\end{IEEEbiography}

\begin{IEEEbiography}[\includegraphics{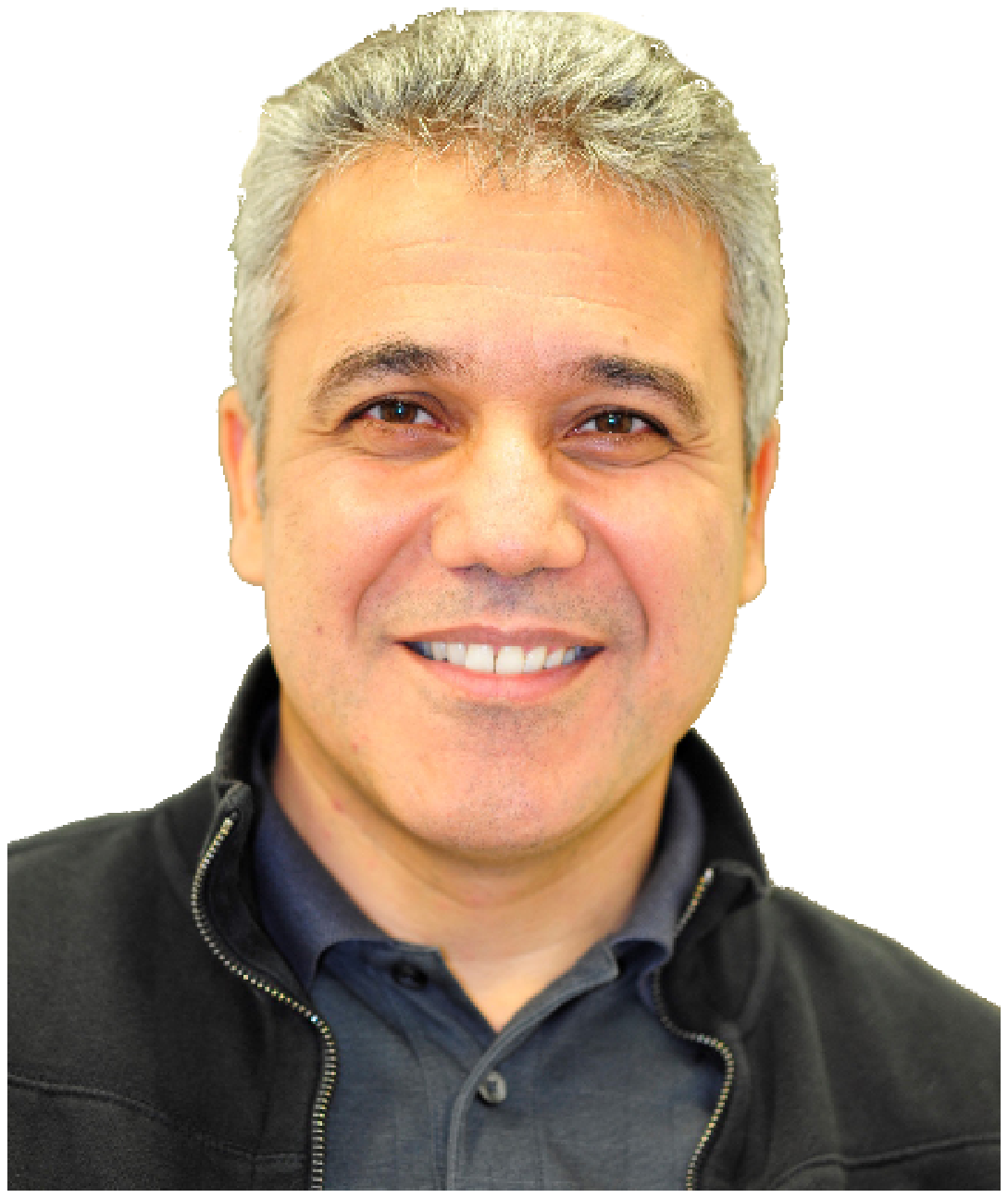}]{Huseyin Arslan}
 (S\textquoteright 95\textendash M\textquoteright 98\textendash SM\textquoteright 04\textendash F\textquoteright 16)
received his B.S. degree from Middle East Technical University, Ankara,
Turkey, in 1992 and his M.S. and PhD. degrees from Southern Methodist
University, Dallas, TX, USA, in 1994 and 1998, respectively. From
January 1998 to August 2002, he was with the research group of Ericsson
Inc., NC, USA, where he was involved with several projects related
to 2G and 3G wireless communication systems. Since August 2002, he
has been with the Department of Electrical Engineering, University
of South Florida, Tampa, FL, USA, where he is a Professor. In December
2013, he joined joined Istanbul Medipol University, Istanbul, Turkey,
where he has worked as the Dean of the School of Engineering and Natural
Sciences. His current research interests include waveform design for
5G and beyond, physical layer security, dynamic spectrum access, cognitive
radio, coexistence issues on heterogeneous networks, aeronautical
(high altitude platform) communications, and \emph{in vivo} channel
modeling and system design. He is currently a member of the editorial
board for the Sensors Journal and the IEEE Surveys and Tutorials.
He is a fellow of the IEEE.
\end{IEEEbiography}

\end{document}